# Cheap Textile Dam Protection of Seaport Cities against Hurricane Storm Surge Waves, Tsunamis, and Other Weather-Related Floods

**Alexander A. Bolonkin**
C & R, 1310 Avenue R, Suite 6-F
Brooklyn, New York 11229, USA
aBolonkin@juno.com, http://Bolonkin.narod.ru

## Abstract

Author offers to complete research on a new method and cheap applicatory design for land and sea textile dams. The offered method for the protection of the USA's major seaport cities against hurricane storm surge waves, tsunamis, and other weather-related inundations is the cheapest (to build and maintain of all extant anti-flood barriers) and it, therefore, has excellent prospective applications for defending coastal cities from natural weather-caused disasters. It may also be a very cheap method for producing a big amount of cyclical renewable hydropower, land reclamation from the ocean, lakes, riverbanks, as well as land transportation connection of islands, and islands to mainland, instead of very costly over-water bridges and underwater tunnels.

## Introduction

In this statement, we consider the protection of important coastal urbanized regions against tropical cyclone (hurricane), tsunami, and other such costly inundations.

**1. A tropical cyclone (hurricane)** is a storm system fueled by the heat released when moist air rises and the water vapor in it condenses. The term describes the storm's origin in the tropics and its cyclonic nature, which means that its circulation is counterclockwise in the northern hemisphere and clockwise in the southern hemisphere. Tropical cyclones are distinguished from other cyclonic windstorms such as nor'easters, European windstorms, and polar lows by the heat mechanism that fuels them, which makes them "warm core" storm systems.

Depending on their location and strength, there are various terms by which tropical cyclones are known, such as hurricane, typhoon, tropical storm, cyclonic storm and tropical depression.

Tropical cyclones can produce extremely strong winds, tornadoes, torrential rain, high waves, and storm surges. The heavy rains and storm surges can produce extensive flooding. Although their effects on human populations can be devastating, tropical cyclones also can have beneficial effects by relieving drought conditions. They carry heat away from the tropics, an important mechanism of the global atmospheric circulation that maintains equilibrium in the earth's troposphere.

An average of 86 tropical cyclones of tropical storm intensity form annually worldwide, with 47 reaching hurricane/typhoon strength, and 20 becoming intense tropical cyclones (at least of Category 3 intensity).

Worldwide, tropical cyclone activity peaks in late summer when water temperatures are warmest. However, each particular basin has its own seasonal patterns. On a worldwide scale, May is the least active month, while September is the most active.[22]

In the North Atlantic, a distinct hurricane season occurs from June 1 to November 30, sharply peaking from late August through September. The statistical peak of the North Atlantic hurricane season is September 10. The Northeast Pacific has a broader period of activity, but in a similar time frame to the Atlantic. The Northwest Pacific sees tropical cyclones year-round, with a minimum in February and a peak in early September. In the North Indian basin, storms are most common from April to December, with peaks in May and November.[22]

Table No. 1.

| Season Lengths and Seasonal Averages | | | | | |
|---|---|---|---|---|---|
| Basin | Season Start | Season End | Tropical Storms (>34 knots) | Tropical Cyclones (>63 knots) | Category 3+ Tropical Cyclones (>95 knots) |
| Northwest Pacific | – | – | 26.7 | 16.9 | 8.5 |
| South Indian | October | May | 20.6 | 10.3 | 4.3 |
| Northeast Pacific | May | November | 16.3 | 9.0 | 4.1 |
| North Atlantic | June | November | 10.6 | 5.9 | 2.0 |
| Australia Southwest Pacific | October | May | 10.6 | 4.8 | 1.9 |
| North Indian | April | December | 5.4 | 2.2 | 0.4 |

A mature tropical cyclone can release heat at a rate upwards of $6 \times 10^{14}$ watts.[3] Tropical cyclones on the open sea cause large waves, heavy rain, and high winds, disrupting international shipping and sometimes sinking ships. However, the most devastating effects of a tropical cyclone occur when they cross coastlines, making landfall. A tropical cyclone moving over land can do direct damage in four ways:

- **High winds** - Hurricane strength winds can damage or destroy vehicles, buildings, bridges, etc. High winds also turn loose debris into flying projectiles, making the outdoor environment even more dangerous.

- **Storm surge** - Tropical cyclones cause an increase in sea level, which can flood coastal communities. This is the worst effect, as historically cyclones claimed 80% of their victims when they first strike shore.
- **Heavy rain** - The thunderstorm activity in a tropical cyclone causes intense rainfall. Rivers and streams flood, roads become impassable, and landslides can occur. Inland areas are particularly vulnerable to freshwater flooding, due to residents not preparing adequately.
- **Tornado activity** - The broad rotation of a hurricane often spawns tornadoes. Also, tornadoes can be spawned as a result of eyewall mesovortices, which persist until landfall. While these tornadoes are normally not as strong as their non-tropical counterparts, they can still cause tremendous damage.[31]

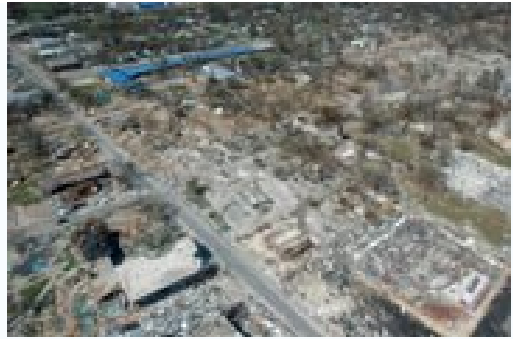

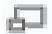

**Fig.1.** The aftermath of Hurricane Katrina in Gulfport, Mississippi. Katrina was the costliest tropical cyclone in United States history.

Often, the secondary effects of a tropical cyclone are equally damaging. These include:

- **Disease** - The wet environment in the aftermath of a tropical cyclone, combined with the destruction of sanitation facilities and a warm tropical climate, can induce epidemics of disease which claim lives long after the storm passes. One of the most common post-hurricane injuries is stepping on a nail in storm debris, leading to a risk of tetanus or other infection. Infections of cuts and bruises can be greatly amplified by wading in sewage-polluted water. Large areas of standing water caused by flooding also contribute to mosquito-borne illnesses.
- **Power outages** - Tropical cyclones often knock out power to tens or hundreds of thousands of people (or occasionally millions if a large urban area is affected), prohibiting vital communication and hampering rescue efforts.
- **Transportation difficulties** - Tropical cyclones often destroy key bridges, overpasses, and roads, complicating efforts to transport food, clean water, and medicine to the areas that need it.

. Hurricane Katrina is the most obvious example, as it devastated the region that had been revitalized after Hurricane Camille. Of course, many former residents and businesses do relocate to inland areas away from the threat of future hurricanes as well.

While the number of storms in the Atlantic has increased since 1995, there seems to be no signs of a numerical global trend; the annual global number of tropical cyclones remains about $90 \pm 10$. However, there is some evidence that the intensity of hurricanes is increasing. "Records of hurricane activity worldwide show an upswing of both the maximum wind speed in and the duration of hurricanes. The energy released by the average hurricane (again considering all hurricanes worldwide) seems to have increased by around 70% in the past 30 years or so, corresponding to about a 15% increase in the maximum wind speed and a 60% increase in storm lifetime."

Atlantic storms are certainly becoming more destructive financially, since five of the ten most expensive storms in United States history have occurred since 1990. This can be attributed to the increased intensity and duration of hurricanes striking North America and to the number of people living in susceptible coastal area following increased development in the region since the last surge in Atlantic hurricane activity in the 1960s.

Tropical cyclones that cause massive destruction are fortunately rare, but when they happen, they can cause damage in the range of billions of dollars and disrupt or end thousands of lives.

The deadliest tropical cyclone on record hit the densely populated Ganges Delta region of Bangladesh on November 13, 1970, likely as a Category 3 tropical cyclone. It killed an estimated 500,000 people. The North Indian basin has historically been the deadliest, with several storms since 1900 killing over 100,000 people, each in Bangladesh.

In the Atlantic basin, at least three storms have killed more than 10,000 people. Hurricane Mitch during the 1998 Atlantic hurricane season caused severe flooding and mudslides in Honduras, killing about 18,000 people and changing the landscape enough that entirely new maps of the country were needed. The Galveston Hurricane of 1900, which made landfall at Galveston, Texas as an estimated Category 4 storm, killed 8,000 to 12,000 people, and remains the deadliest natural disaster in the history of the United States. The deadliest Atlantic storm on record was the Great Hurricane of 1780, which killed about 22,000 people in the Antilles.

Hurricane Iniki in 1992 was the most powerful storm to strike Hawaii in recorded history, hitting Kauai as a Category 4 hurricane, killing six and causing $3 billion in damage. Other destructive Pacific hurricanes include Pauline and Kenna.

On March 26, 2004, Cyclone Catarina became the first recorded South Atlantic cyclone (cyclone is the southern hemispheric term for *hurricane*). Previous South Atlantic cyclones in 1991 and 2004 reached only tropical storm strength. Tropical cyclones may have formed there before 1960 but were not observed until weather satellites began monitoring the Earth's oceans in that year.

A tropical cyclone need not be particularly strong to cause memorable damage; Tropical Storm Thelma, in November 1991 killed thousands in the Philippines even though it never became a typhoon; the damage from Thelma was mostly due to flooding, not winds or storm surge. In 1982, the unnamed tropical depression that eventually became Hurricane Paul caused the deaths of around 1,000 people in Central America due to the effects of its rainfall. In addition, Hurricane Jeanne in 2004 caused the majority of its damage in Haiti, including approximately 3,000 deaths, while just a tropical depression.

On August 29, 2005, Hurricane Katrina made landfall in Louisiana and Mississippi. The U.S. National Hurricane Center, in its August review of the tropical storm season stated that Katrina was probably the worst natural disaster in U.S. history. Currently, its death toll is at least 1,836, mainly from flooding and the aftermath in New Orleans, Louisiana and the Mississippi Gulf Coast. It is also estimated to have caused $81.2 billion in property damage. Before Katrina, the costliest system in monetary terms had been 1992's Hurricane Andrew, which caused an estimated $39 billion (2005 USD) in damage in Florida.

2. A **tsunami** is a series of waves when a body of water, such as an ocean is rapidly displaced on a massive scale. Earthquakes, mass movements above or below water, volcanic eruptions and other underwater explosions, landslides and large meteorite impacts all have the potential to generate a tsunami. The effects of a tsunami can range from unnoticeable to devastating. Tsunamis are common throughout Japanese history, as 195 events in Japan have been recorded.

A tsunami has a much smaller amplitude (wave heights) offshore, and a very long wavelength (often hundreds of kilometres long), which is why they generally pass unnoticed at sea, forming only a passing "hump" in the ocean.

Tsunamis can be generated when the sea floor abruptly deforms and vertically displaces the overlying water. Such large vertical movements of the Earth's crust can occur at plate boundaries. Subduction earthquakes are particularly effective in generating tsunamis. As an Oceanic Plate is subducted beneath a Continental Plate, it sometimes brings down the lip of the Continental with it. Eventually, too much stress is put on the lip and it snaps back, sending shockwaves through the Earth's crust, causing a tremor under the sea, known as an Undersea Earthquake.

Sub-marine landslides (which are sometimes triggered by large earthquakes) as well as collapses of volcanic edifices may also disturb the overlying water column as sediment and rocks slide downslope and are redistributed across the sea floor. Similarly, a violent submarine volcanic eruption can uplift the water column and form a tsunami.

Tsunamis are surface gravity waves that are formed as the displaced water mass moves under the influence of gravity and radiate across the ocean like ripples on a pond.

In the 1950s it was discovered that larger tsunamis than previously believed possible could be caused by landslides, explosive volcanic action and impact events. These phenomena rapidly displace large volumes of water, as energy from falling debris or expansion is transferred to the water into which the debris falls. Tsunamis caused by these mechanisms, unlike the ocean-wide tsunamis caused by some earthquakes, generally dissipate quickly and rarely affect coastlines distant from the source due to the small area of sea affected. These events can give rise to much larger local shock waves (solitons), such as the landslide at the head of Lituya Bay which produced a water wave estimated at 50 – 150 m and reached 524 m up local mountains. However, an extremely large landslide could generate a megatsunami that might have ocean-wide impacts.

While it is not possible to prevent a tsunami, in some particularly tsunami-prone countries some measures have been taken to reduce the damage caused on shore. Japan has implemented an extensive programme of building tsunami walls of up to 4.5 m (13.5 ft) high in front of populated coastal areas. Other localities have built floodgates and channels to redirect the water from incoming tsunamis. However, their effectiveness has been questioned, as tsunamis are often higher than the barriers. For instance, the tsunami which hit the island of Hokkaido on July 12, 1993 created waves as much as 30 m (100 ft) tall - as high as a 10-story building. The port town of Aonae was completely surrounded by a tsunami wall, but the waves washed right over the wall and destroyed all the wood-framed structures in the area. The wall may have succeeded in slowing down and moderating the height of the tsunami but it did not prevent major destruction and loss of life.

Japan is a nation with the most recorded tsunamis in the world. The earliest recorded disaster being that of the 684 A.D. Hakuho Quake. The number of tsunamis in Japan totals 195 over a 1,313 year period, averaging one event every 6.7 years, the highest rate of occurrence in the world. These waves have hit with such violent fury that entire towns have been destroyed. In 1896 Sanriku, Japan, with a population of 20,000, suffered such a devastating fate.

On December 26, 2004, an undersea earthquake measuring 9.0 on the Richter scale occurred 160 km (100 mi) off the western coast of Sumatra, Indonesia. It was the fifth largest earthquake in recorded history and generated massive tsunamis, which caused widespread devastation when they hit land, leaving an estimated 250,000 people dead in countries around the Indian Ocean.

The 2004 Indian Ocean earthquake, which had a magnitude of 9.3, triggered a series of lethal tsunamis on December 26, 2004 that **killed approximately 230,000 people** (including 168,000 in Indonesia alone), making it the deadliest tsunami in recorded history. The tsunami killed people over an area ranging from the immediate vicinity of the quake in Indonesia, Thailand and the north-western coast of Malaysia to thousands of kilometres away in Bangladesh, India, Sri Lanka, the Maldives, and even as far as Somalia, Kenya and Tanzania in eastern Africa.

Unlike in the Pacific Ocean, there was no organized alert service covering the Indian Ocean. This was in part due to the absence of major tsunami events since 1883 (the Krakatoa eruption, which killed 36,000 people). In light of the 2004 Indian Ocean tsunami, UNESCO and other world bodies have called for a global tsunami monitoring system.

3) A **flood (inundation)** is an overflow of water, an expanse of water submerging land, a deluge. In the sense of "flowing water", the word is applied to the inflow of the tide, as opposed to the outflow or "ebb". *The* Flood, the great Universal Deluge of myth and perhaps of history is treated at Deluge in mythology.

Since prehistoric times people have lived by the seas and rivers for the access to cheap and quick transportation and access to food sources and trade; without human populations near natural bodies of water, there would be no concern for floods. However fertile soil in a river delta is subject to regular inundation from normal variation in precipitation.

Floods from the sea can cause overflow or overtopping of flood-defenses like dikes as well as flattening of dunes or bluffs. Land behind the coastal defence may be inundated or experience damage. A flood from sea may be caused by a heavy storm (storm surge), a high tide, a tsunami, or a combination thereof. As many urban communities are located near the coast this is a major threat around the world.

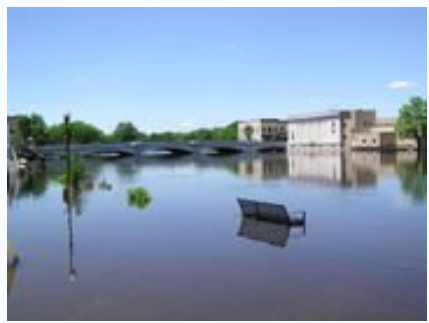

**Fig. 2.** Rock River floodwaters in downtown Fort Atkinson, Wisconsin.

Many rivers that flow over relatively flat land border on broad flood plains. When heavy the deposition of silt on the rich farmlands and can result in their eventual depletion. The annual cycle of flood and farming was of great significance to many early farming cultures, most famously to the ancient Egyptians of the Nile river and to the Mesopotamians of the Tigris and Euphrates rivers .

A flood happens when an area of land, usually low-lying, is covered with water. The worst floods usually occur when a river overflows its banks. An example of this is the January 1999 Queensland floods, which swamped south-eastern Queensland. Floods happen when soil and vegetation cannot absorb all the water. The water then runs off the land in quantities that cannot be carried in stream channels or kept in natural ponds or man-made reservoirs.

Periodic floods occur naturally on many rivers, forming an area known as the flood plain. These river floods usually result from heavy rain, sometimes combined with melting snow, which causes the rivers to overflow their banks. A flood that rises and falls rapidly with little or no advance warning is called a flash flood. Flash floods usually result from intense rainfall over a relatively small area. Coastal areas are occasionally flooded by high tides caused by severe winds on ocean surfaces, or by tidal waves caused by undersea earthquakes. There are often many causes for a flood.

Monsoon rainfalls can cause disastrous flooding in some equatorial countries, such as Bangladesh, Hurricanes have a number of different features which, together, can cause devastating flooding. One is the storm surge (sea flooding as much as 8 metres high) caused by the leading edge of the hurricane when it moves from sea to land. Another is the large amounts of precipitation associated with hurricanes. The eye of a hurricane has extremely low pressure, so sea level may rise a few metres in the eye of the storm. This type of coastal flooding occurs regularly in Bangladesh.

In Europe floods from sea may occur as a result from heavy Atlantic storms, pushing the water to the coast. Especially in combination with high tide this can be damaging.

Under some rare conditions associated with heat waves, flash floods from quickly melting mountain snow have caused loss of property and life.

Undersea earthquakes, eruptions of island volcanos that form a caldera, (such as Thera or Krakatau) and marine landslips on continental shelves may all engender a tidal wave called a tsunami that causes destruction to coastal areas. See the *tsunami* article for full details of these marine floods.

Floods are the most frequent type of disaster worldwide. Thus, it is often difficult or impossible to obtain insurance policies which cover destruction of property due to flooding, since floods are a relatively predictable risk.

- In 1983 the Pacific Northwest saw one of their worst winter floods. And the some of the Northwest states saw their wettest winter yet. The damage was estimated at 1.1 billion dollars.* In 1965 Hurricane Betsy flooded large areas of New Orleans for up to 10 days, drowning around 40 people.
- In 1957, storm surge flooding from Hurricane Audrey killed about 400 people in southwest Louisiana.
- The Hunter Valley floods of 1955 in New South Wales destroyed over 100 homes and caused 45,000 to be evacuated.
- The North Sea Flood of 1953 caused over 2,000 deaths in the Dutch province of Zeeland and the United Kingdom and led to the construction of the Delta Works and the Thames Barrier.
- The Lynmouth flood of 1952 killed only 34 people, however it was very destructive and destroyed over 80 buildings.
- The 1931 Huang He flood caused between 800,000 and 4,000,000 deaths in China, one of a series of disastrous floods on the Huang He.

- The Great Mississippi Flood in 1927 was one of the most destructive floods in United States history.

The 2005 tragedy of New Orleans shows that disregard of protection of the USA's coastal cities (New York, Los Angles-San Pedro) from strong storm-caused waves, hurricane storm surges, and small tsunamis gives rise to gigantic damages, material losses, human deaths and injuries.

The Metropolitan East Coast (MEC) region -- with New York City at its center -- has nearly 20 million people, a $1 trillion economy, and $2 trillion worth of built assets, nearly half of which are complex infrastructure.

Many elements of transportation and other essential infrastructure systems in the MEC region, and even some of its regular building stock, are located at elevations from 6 to 20 feet above current sea level. This is well within the range of expected coastal storm surge elevation of 8 to more than 20 feet for tropical (hurricanes) and extra-tropical storms. Depending on which climate change scenarios apply, the sea level regional rise over the next 100 years will accelerate and amount to at most 3 feet by the year 2100. This seemingly modest increase in sea level has the effect to raise the frequency of coastal storm surges and related flooding by factors of 2 to 10, with an average of about 3.

The rate of financial losses incurred from these coastal floods will increase accordingly. Expected annualized losses from coastal storms, already on the order of about $1 billion per year, would be small enough to be absorbed by the $1 trillion economy of the region. However, actual losses do not occur in regular annualized doses. Rather, they occur during infrequent extreme events that can cause losses of hundreds of billions of dollars for the largest events, albeit with low probability.

| **Ten deadliest natural disasters** | | | Table 2 |
|---|---|---|---|
| *Rank Event* | *Location* | *Date* | *Death Toll (Estimate)* |
| 1. 1931 Yellow River flood | Yellow River, China | Summer 1931 | 850,000-4,000,000 |
| 2. 1887 Yellow River flood | Yellow River, China | September-October 1887 | 900,000-2,000,000 |
| 3. 1970 Bhola cyclone | Ganges Delta, East Pakistan | November 13, 1970 | 500,000-1,000,000 |
| 4. Earthquake | Eastern Mediterranean | 1201 | 1,000,000 |
| 5. 1938 Yellow River flood | Yellow River, China | June 9th, 1938 | 500,000 - 900,000 |
| 6. Shaanxi Earthquake | Shaanxi Province, China | January 23, 1556 | 830,000 |
| 7. 2004 Indian Ocean earthquake/tsunami | Indian Ocean | December 26, 2004 | 225,000-275,000 |
| 8. Tropical Cyclone | Haiphong, Vietnam | 1881 | 300,000 |

| 9. | Flood | Kaifeng, Henan Province, China | 1642 | 300,000 |
| 10. | Earthquake | Tangshan, China | July 28, 1976 | 242,000* |

\* Official Government figure. Estimated death toll as high as 655,000.

## Description of innovation

Current coast-protection dams are built from solid material (heaped stones, concrete, piled soil). They are expensive to emplace and, sometimes, are unsightly. Such dams require detailed on-site research of the surface and sub-surface environment, costly construction and high-quality building efforts over a long period of time (years). Naturally, the coast city inhabitants lose the beautiful sea view and ship passengers are unable to admire the city panorama of the partly hidden city (New Orleans, which is below sea level). The sea coastal usually has a complex geomorphic configuration that greatly increases the length and cost of dam protections.

Authors offer to protect seaport cities against hurricane storm surge waves, tsunamis, and other weather related-coastal and river inundations by new special design of the water and land textile dams.

The offered dam is shown in Fig. 3a below. One contains the floats 4, textile (thin film) 3 and support cables 5. The textile (film) is connected a top edge to the floats, the lower edge to a sea bottom. In calm weather the floats are located on the sea surface (Fig. 1a) or at the sea bottom (Fig. 1b). In stormy weather, hurricane, predicted tsunami the floats automatically raise to top of wave and defend the city from any rapid increase of seawater level (Fig. 3c).

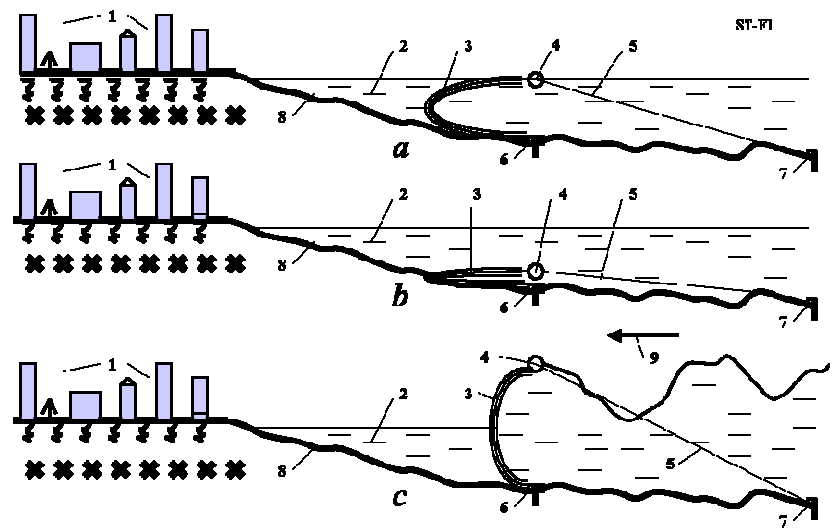

**Fig. 3.** Protection city against hurricane storm surge waves, tsunamis, and other weather related-coastal inundations by textile (film) membrane located in sea (ocean). (*a*) - position of membrane on a sea surface in a calm weather; (*b*) - position of membrane on a sea bottom in a calm weather; (*c*) - position of membrane in hurricane storm surge waves, tsunamis, and other weather related-coastal inundations. Notations: 1 - city, 2 - sea (ocean), 3 - membrane, 4 - float, 5 - support cable, 6 - connection of membrane to a sea bottom, 7 - connection of support cable to a sea bottom, 8 - sea bottom, 9 - wind.

This textile-based dam's cost-to-build is thousands of times cheaper then a massive concrete dam, and a textile infrastructure may be assembled in few months instead of years! They may be installed on ground surface around vital or important infrastructure objects (entries to subway tunnels, electricity power plants, civic airport, and so on) or around a high-value part of the city (example, Manhattan Island) if inundation poses a threat to the city (Fig. 4). These textile protections are mobile and can be relocated and installed in few days if hurricane is predictably moving to given city. They can defend the noted object or city from stormy weather inundation, tsunamis, and large waves of height up 30 and more feet.

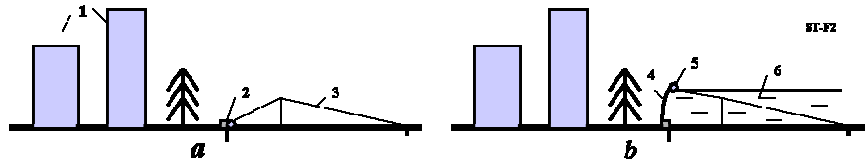

**Fig.4.** Protection city against hurricane storm surge waves, tsunamis, and other weather related-coastal inundations by textile (film) membrane located on ground surface. (*a*) - position of membrane on a ground surface in a compact form in a calm weather; (*b*) - position of membrane in hurricane storm surge waves, tsunamis, and other weather related-coastal inundations. Notations: 1 - city, 2 - membrane in the compact form, 3 - support cable, 4 - membrane, 5 - float, 6 -water, 7 - connection of support cable to a sea bottom.

The offered textile dam may be also used as a big source of electricity. They can be built as the dams in rivers and it is used as water dams for the electric station (Fig. 5).

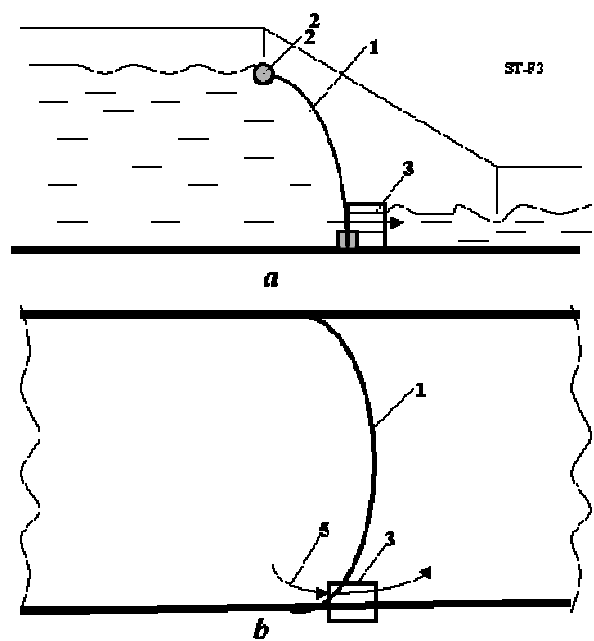

**Fig. 5.** Textile dam and electric station in a river. (*a*) side view; (*b*) - top view. Notations: 1 - textile dam, 2 - float, 3 - electric station, 4 - water flow.

They also can be used as the dams for an ebb - flow sea electric station (Fig. 6).

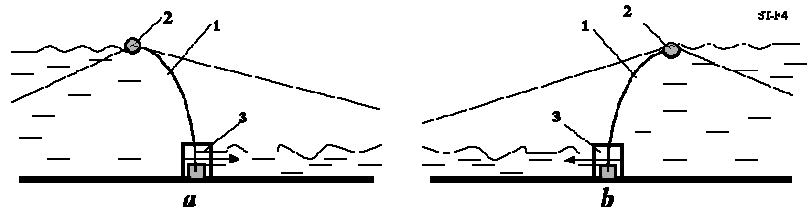

**Fig.6**. Tidal (ebb and flow) electric station with textile dam. (*a*) - ebb; (*b*) - flow. Notations: 1 -textile membrane, 2 - float, 3 - electric station.

Double textile dams can be also used for drying a big area of a shallow sea and converting its to industrial and farmland zones or for connection of closely islands (Fig. 7). It may be cheaper then to build an expensive bridges or underground tunnels. For security the textile dams must be double (Fig. 7) and area located lower a sea level must be divided in zones separated by additional membranes.

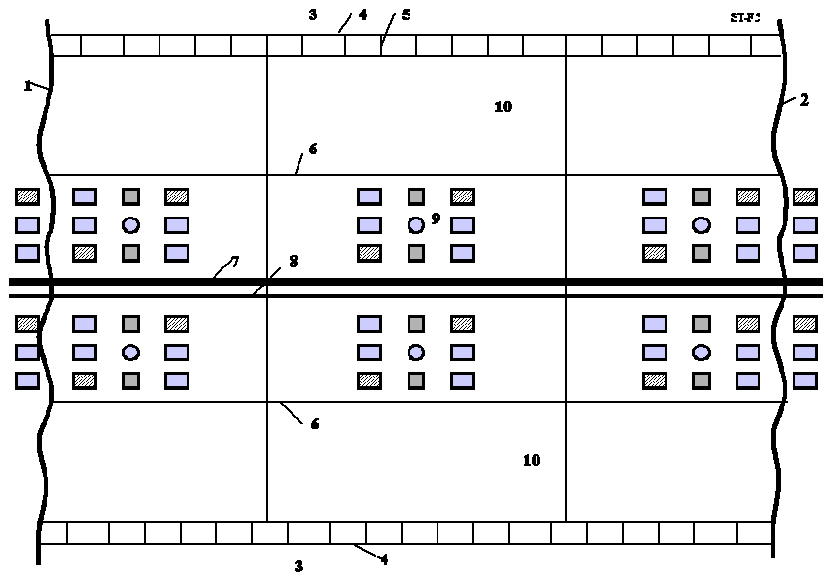

**Fig.7**. Ground connection between islands or converting a shallow sea in dry land by textile membranes. Notations: 1 - the fist island, 2 - the second island, 3 - sea, 4 - double textile membrane, 5 - textile partitions between the double membranes, 6 - additional emergency ground textile partition, 7 - car (track) highway, 8 - railroad, 9 - dwelling (industrial) zone; 10 - farmland.

The membranes must be made from artificial fiber or a film. The many current artificial fibers are cheap, have very high safety tensile stress (some times more the steel!) and chemical stability. They can work as dam some tens years. They are easy for repair.

### Theory and Computation

1. **Force** $P$ [N/m$^2$] for 1 m$^2$ of dam is

$$P = g\gamma h$$
, (1)

where $g$ = 9.81 m/s2 is the Earth gravity; $\gamma$ is water density, $\gamma$ =1000 kg/m3; $h$ is difference between top and lower levels of water surfaces, m (see computation in Fig. 8).

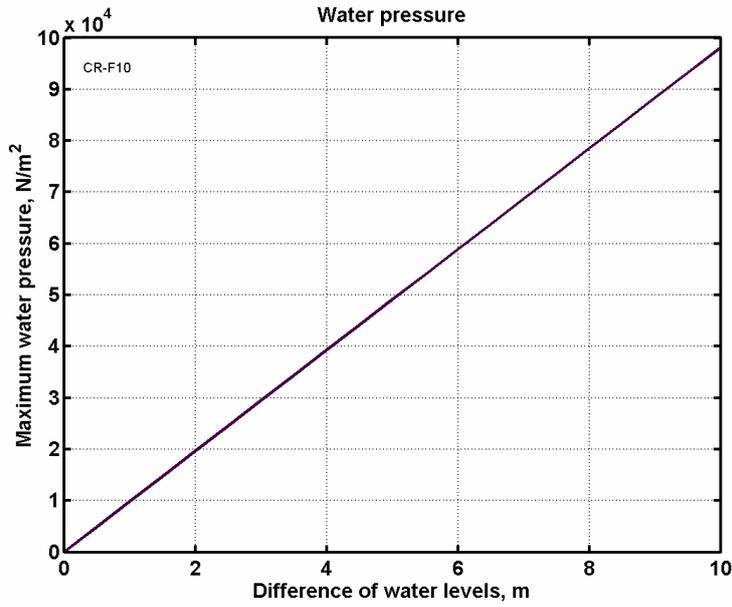

**Fig. 8**. Water pressure via difference of water levels

## 2. Water power $N$ [W] is

$$N = \eta g m h, \quad m = \gamma v S, \quad v = \sqrt{2gh}, \quad N = \eta g \gamma h S \sqrt{2gh}, \quad N/S \approx 43.453 \eta h^{1.5}, \quad [\text{kW/m}^2] \qquad (2)$$

where $m$ is mass flow across 1 m width kg/m; $v$ is water speed, m/s; $S$ is turbine area, m²; $\eta$ is coefficient efficiency of the water turbine, $N/S$ is specific power of water turbine, kW/m². Computation is presented in Fig. 9.

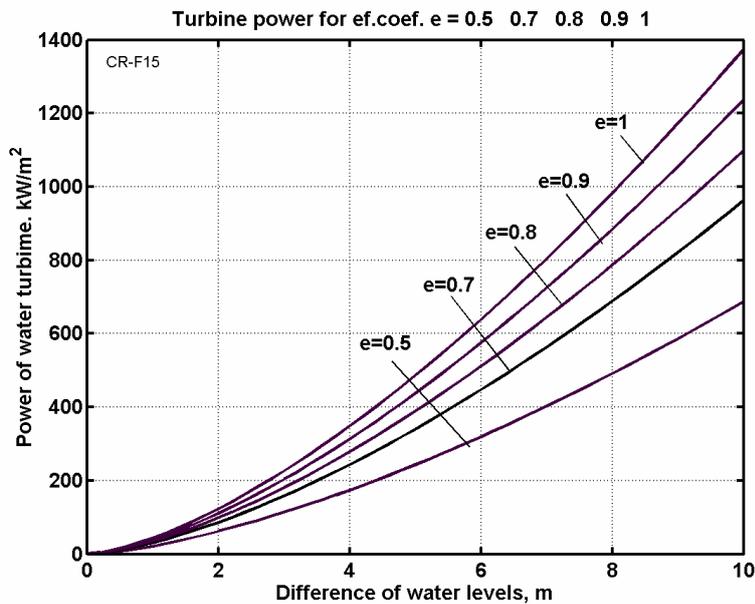

**Fig. 9**. Specific power of a water turbine via difference of water levels and turbine efficiency coefficient

### 3. Film thickness is

$$\delta = \frac{g\gamma\,h^2}{2\sigma},$$ (3)

where σ is safety film tensile stress, N/m$^2$. Results of computation are in Fig. 10. The fibrous material (Fiber B, PRD-49) has $\sigma = 312$ kg/mm$^2$ and specific gravity $\gamma = 1.5$ g/cm$^3$ [7].

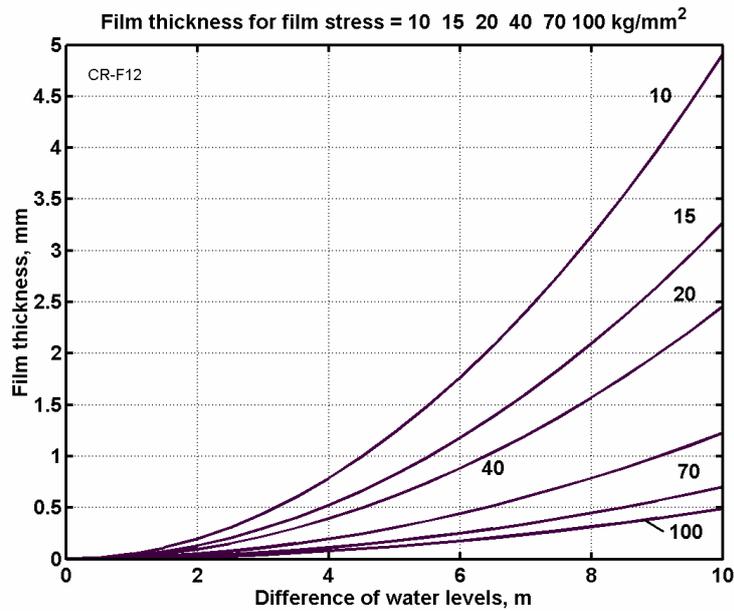

**Fig. 10**. Film (textile) thickness via difference of water levels safety film (textile) tensile stress.

### 4. The film weight of 1m width is

$$W_f = 1.2\delta\gamma H,$$ (4)

Computation are in Fig. 11. If our dam has long $L$ m, we must multiple this results by $L$.

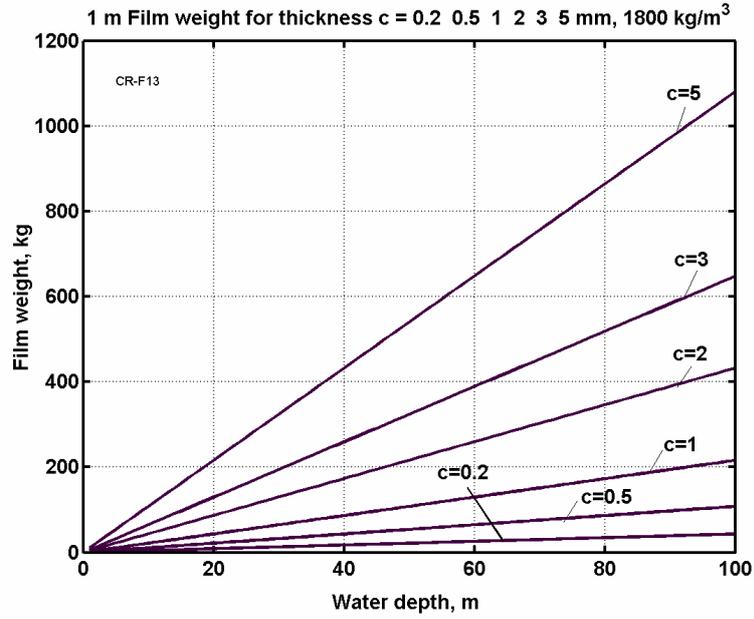

**Fig. 11**. 1 m Film weight via the deep of dam and film thickness $c$, density 1800 kg/m$^3$ .

**5. The diameter $d$ of the support cable is**

$$T = \frac{Pl_2}{2}, \quad S = \frac{T}{\sigma}, \quad d = \sqrt{\frac{4S}{\pi}},$$

(5)

where $T$ is cable force, N; $l_2$ is distance between cable, m; $S$ is cross-section area, m$^2$.

Computation is presented in fig. 12. The total weight of support cable is

$$W_c \approx 2\gamma_c HSL / l_2, \quad W_a = \gamma_c SL,$$

(6)

where $\gamma_c$ is cable density, kg/m$^3$; $L$ is length of dam, m; $W_a$ is additional

(connection of banks) cable, m. The cheap current fiber has $\sigma = 620$ kg/mm$^2$ and specific gravity $\gamma = 1.8$ g/cm$^3$ [7].

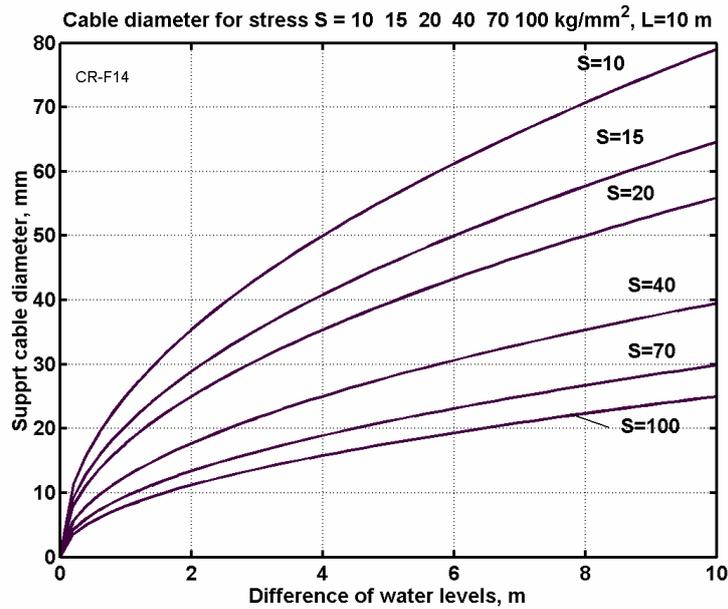

**Fig. 12**. Diameter of the support cable via difference of a water levels and the safety tensile stress for every 10 m textile dam .

**6. Maximum sea raise of water** from hurricane versus wind speed is

$$h = \frac{\rho V^2}{2\gamma}, \qquad\qquad (7)$$

where $h$ is water raising, m; $\rho = 1.225$ kg/m³; $V$ is wind speed, m/s; $\gamma = 1000$ kg/m³ is water density. Computation is presented in Fig. 13.

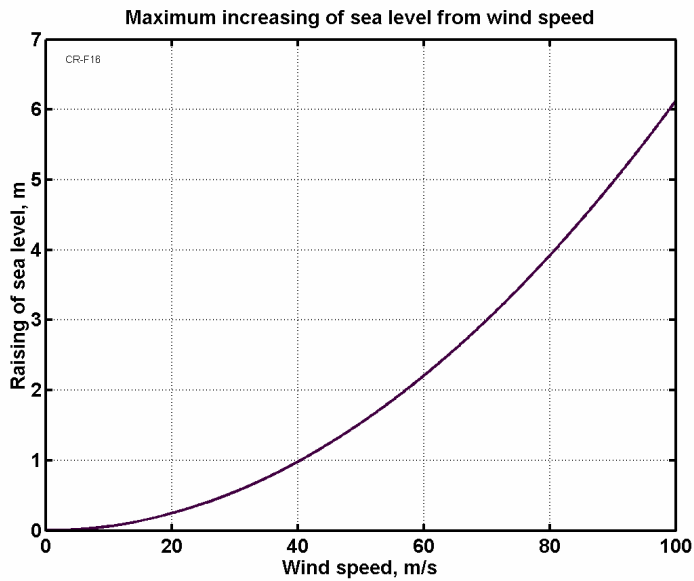

**Fig.13**. Raising of sea level via wind speed.

Wind speed is main magnitude which influences in the water raising. The direction of wind, rain, general atmospheric pressure, deep, and relief of sea bottom, Moon phase, also influence to the water raising and can decreases or increases the local sea level computed by Equation (7). For example, in hurricane "eye" the wind is absent, but atmospheric pressure is very low and sea level is high.

## Application

Using the graphs above, we can estimate the relevant physical parameters of many interesting macroprojects [1] - [7].

## Summary

Author offered and researched the new method and cheap design the land and sea textile (film) dams. The offered method of the protection of seaport cities against hurricane storm surge waves, tsunamis, and other weather related inundations is cheapest and has the very perspective applications for defense from natural weather disasters. That is also method for producing a big amount of renewable cheap energy, getting a new land for sea (and non-sea) countries. However, there are important details not considered in this research. It is recommended the consulting with author for application this protection.